# First Evidence for Direct $CP$ Violation in Beauty to Charmonium Decays

R. Aaij *et al.*[*]
(LHCb Collaboration)

The $CP$ asymmetry and branching fraction of the Cabibbo-Kobayashi-Maskawa-suppressed decay $B^+ \to J/\psi\pi^+$ are precisely measured relative to the favored decay $B^+ \to J/\psi K^+$ using a sample of proton-proton collision data corresponding to an integrated luminosity of 5.4 $\mathrm{fb}^{-1}$ recorded at a center-of-mass energy of 13 TeV during 2016–2018. The results of the $CP$ asymmetry difference and branching fraction ratio are $\Delta\mathcal{A}^{CP} \equiv \mathcal{A}^{CP}(B^+ \to J/\psi\pi^+) - \mathcal{A}^{CP}(B^+ \to J/\psi K^+) = (1.29 \pm 0.49 \pm 0.08) \times 10^{-2}$, $\mathcal{R}_{\pi/K} \equiv [\mathcal{B}(B^+ \to J/\psi\pi^+)/\mathcal{B}(B^+ \to J/\psi K^+)] = (3.852 \pm 0.022 \pm 0.018) \times 10^{-2}$, where the first uncertainties are statistical and the second are systematic. A combination with previous LHCb results based on data collected at 7 and 8 TeV in 2011 and 2012 yields $\Delta\mathcal{A}^{CP} = (1.42 \pm 0.43 \pm 0.08) \times 10^{-2}$ and $\mathcal{R}_{\pi/K} = (3.846 \pm 0.018 \pm 0.018) \times 10^{-2}$. The combined $\Delta\mathcal{A}^{CP}$ value deviates from zero by 3.2 standard deviations, providing the first evidence for direct $CP$ violation in the amplitudes of beauty decays to charmonium final states.

Violation of the charge-parity ($CP$) symmetry is one of the conditions necessary to generate the matter-antimatter asymmetry in the Universe [1]. Beauty decays to charmonium final states, governed by $b \to c\bar{c}q$ quark-level transitions (where $q = s, d$), play a pivotal role in the study of $CP$ violation. In general, $CP$ violation can arise directly from the interference of the leading-order $W$-emission (tree) amplitude and the loop (penguin) amplitudes of such decays, manifesting as a small decay-rate asymmetry between two $CP$-conjugated processes, referred to as direct $CP$ violation. For neutral $B$ mesons, $CP$ violation can also arise from the interference of the direct decay and the decay after flavor mixing, manifesting as a time-dependent decay-rate asymmetry. Precision measurements of the weak phases $2\beta = 2\phi_1 \equiv 2\arg[-(V_{cd}V_{cb}^*)/(V_{td}V_{tb}^*)]$ and $2\beta_s \equiv 2\arg[-(V_{ts}V_{tb}^*)/(V_{cs}V_{cb}^*)]$, where $V_{ij}$ are elements of the Cabibbo-Kobayashi-Maskawa (CKM) matrix [2,3], from the time-dependent $CP$ asymmetries in the golden channels $B^0 \to J/\psi K^0$ [4–6] and $B_s^0 \to J/\psi K^+K^-$ [7–10], respectively, have provided stringent tests of the standard model (SM). An open issue in the $2\beta_{(s)}$ determination is related to the effects of the subleading contributions from highly suppressed penguin diagrams in $b \to c\bar{c}s$ transitions, which need to be fully understood for more precise tests of the SM but are difficult to calculate reliably in theory. Such effects can eventually be controlled with measurements of penguin-enhanced $b \to c\bar{c}d$ transitions, such as $B^+ \to J/\psi\pi^+$ decays [11], as detailed in Refs. [12–15]. Specifically, measurements of the direct $CP$ violation and decay rate of the $B^+ \to J/\psi\pi^+$ process, together with time-dependent $CP$ asymmetries measured in both $B^0 \to J/\psi\pi^0$ and $B_s^0 \to J/\psi\bar{K}^0$ decays, allow the penguin effects in $B^0 \to J/\psi K^0$ to be included in the determination of the phase $2\beta$ using approximate SU(3) flavor symmetry [12–15].

Unlike the case of $b \to c\bar{c}s$ transitions, the penguin contributions in $B^+ \to J/\psi\pi^+$ are not CKM suppressed with respect to the leading-order tree diagram. Thus, sizable direct $CP$ violation up to the percent level could arise from interference between the tree and penguin contributions [16,17], which is within reach of the LHCb experiment, though unobserved to date. In order to subtract the small difference between the production cross sections of $B^-$ and $B^+$ mesons, the asymmetry is measured with respect to that of the $B^+ \to J/\psi K^+$ decay, where direct $CP$ violation is expected to be negligible due to the dominance of the tree diagram, namely,

$$\Delta\mathcal{A}^{CP} \equiv \mathcal{A}^{CP}(B^+ \to J/\psi\pi^+) - \mathcal{A}^{CP}(B^+ \to J/\psi K^+). \quad (1)$$

Here $\mathcal{A}^{CP}$ is the decay-rate asymmetry between the $B^-$ and $B^+$ mesons. In addition, information on the penguin contributions can be obtained from the ratio of branching fractions [18],

$$\mathcal{R}_{\pi/K} \equiv \frac{\mathcal{B}(B^+ \to J/\psi\pi^+)}{\mathcal{B}(B^+ \to J/\psi K^+)}, \quad (2)$$

[*]Full author list given at the end of the Letter.

where the systematic uncertainties related to the trigger, reconstruction, and selection efficiencies largely cancel out. The LHCb Collaboration has previously measured $\Delta\mathcal{A}^{CP} = [1.82 \pm 0.86(\text{stat}) \pm 0.14(\text{syst})] \times 10^{-2}$, which is consistent with $CP$ conservation, and $\mathcal{R}_{\pi/K} = [3.83 \pm 0.03(\text{stat}) \pm 0.03(\text{syst})] \times 10^{-2}$ [19], using proton-proton ($pp$) collision data collected at center-of-mass energies of 7 and 8 TeV from 2011 to 2012 (Run 1), corresponding to an integrated luminosity of 3 $\text{fb}^{-1}$.

Many efforts have also been made to search for direct $CP$ violation in other $b \to c\bar{c}d$ processes, such as $B^0 \to J/\psi\pi^0$ [20,21], $B_s^0 \to J/\psi\bar{K}^0$ [22], $B^0 \to J/\psi\rho^0$ [23,24], $B_s^0 \to J/\psi\bar{K}^{*0}$ [25], $B^+ \to \psi(2S)\pi^+$ [26,27], $B^+ \to J/\psi\rho^+$ [24,28], $B^+ \to \chi_{c1}(1P)\pi^+$ [29], and $\Lambda_b^0 \to J/\psi p\pi^-$ [30] decays. However, owing to the limited sensitivity, no evidence for direct $CP$ violation has been found in beauty hadron to charmonium decays so far. This Letter presents updated measurements of $\Delta\mathcal{A}^{CP}$ and $\mathcal{R}_{\pi/K}$ using data recorded by LHCb at 13 TeV in 2016–2018 (Run 2), corresponding to an integrated luminosity of 6 $\text{fb}^{-1}$.

The LHCb detector is a single-arm forward spectrometer covering the pseudorapidity range $2 < \eta < 5$, which is described in detail in Refs. [31,32]. The magnetic-field polarity is reversed periodically during data taking to mitigate the differences of reconstruction efficiencies of particles with opposite charges. Datasets corresponding to about half of the total integrated luminosity are recorded with each magnetic-field configuration.

Samples of simulated events are used to study the properties of the signal mode $B^+ \to J/\psi(\to \mu^+\mu^-)\pi^+$ and the control mode $B^+ \to J/\psi(\to \mu^+\mu^-)K^+$. These simulated events are produced with the software described in Refs. [33–37]. The momentum and transverse momentum ($p_\text{T}$) spectra of the $B^+$ mesons as well as the track multiplicity in simulation are corrected to match those in the data. Additionally, the particle identification (PID) performance of the simulation is also calibrated to match that in data evaluated with large control samples [38,39]. The corrections are determined in the initial phase of the analysis and are included in all subsequent steps.

The online event selection used in this study is performed via a trigger system [39] consisting of a hardware stage that selects events containing at least one muon candidate, and two software trigger stages in which events with two tracks identified as muons with $p_\text{T} > 500\ \text{MeV}/c$ are selected. The muon pair is required to have an invariant mass within $\pm 150\ \text{MeV}/c^2$ of the known $J/\psi$ mass [40].

In the off-line selection, the $B^+ \to J/\psi h^+$ candidates (where $h = \pi$, $K$) are formed by combining a $J/\psi$ with a hadron candidate with $p_\text{T}$ above 1 $\text{GeV}/c$. The selection criteria for the $B^+ \to J/\psi\pi^+$ and $B^+ \to J/\psi K^+$ decays are similar except for those related to the identification of the kaon and pion hadrons in the final states. The accompanying hadron is mutually exclusively identified as a pion or kaon using information from the ring-imaging Cherenkov detectors [41] and is required to be inconsistent with originating from any primary $pp$ collision vertex (PV) and consistent with originating from the $J/\psi$ decay vertex. The particle identification criteria achieve a signal efficiency of 96% (92%) for the $B^+ \to J/\psi\pi^+$ ($B^+ \to J/\psi K^+$) decay, while rejecting 97% (99%) of the misidentified cross-feed background coming from the $B^+ \to J/\psi K^+$ ($B^+ \to J/\psi\pi^+$) decay.

Each $B^+$ candidate must be consistent with originating from a PV. A kinematic fit [42] to the signal decay where the $J/\psi$ mass is constrained to its known value [40] is performed to achieve a better resolution of the reconstructed $B$ mass. The remaining $B^+$ candidates with $\cos\theta_h < 0$ are rejected to ensure a clear separation of the $B^+ \to J/\psi\pi^+$ and $B^+ \to J/\psi K^+$ mass peaks in the $J/\psi\pi^+$ mass spectrum, where $\theta_h$ is the angle between the momentum of the accompanying hadron in the $B^+$ rest frame and the $B^+$ momentum in the laboratory frame. Fiducial-volume requirements are also imposed to exclude those $B^+$ candidates with accompanying hadrons at the boundaries of the detector acceptance, where the detection asymmetry is particularly large. Such requirements retain more than 95% of the $B^+ \to J/\psi h^+$ signals.

In order to further suppress the background from random combinations of tracks (combinatorial background), a boosted decision tree (BDT) classifier [43,44] is trained for each of the $B^+ \to J/\psi\pi^+$ and $B^+ \to J/\psi K^+$ decay modes and each year of data taking. The BDT classifier is trained using simulated $B^+ \to J/\psi h^+$ decays as a signal proxy and a sample of data candidates in the upper mass sideband $[5500, 5700]\ \text{MeV}/c$ above the fit range as a background proxy. The kinematic and geometrical variables used as inputs to the BDT classifier include measures of the likelihood that the $h^+$, $\mu^\pm$, $J/\psi$, or $B^+$ candidate comes from the PV; transverse momentum of the $h^+$, $J/\psi$, and $B^+$ candidates; and flight distance and vertex fit quality of the $B^+$ candidate. The $B^+$ candidates with a BDT response below a certain threshold are rejected. This threshold for the $B^+ \to J/\psi\pi^+$ mode is chosen to optimize the signal significance. For $B^+ \to J/\psi K^+$ decays, the threshold is chosen to obtain the same BDT selection efficiency as for the $B^+ \to J/\psi\pi^+$ mode. The optimized BDT selection retains about 95% of both signals while rejecting more than 90% of the combinatorial backgrounds.

An unbinned extended maximum-likelihood fit is performed simultaneously with the mass distributions of the selected $B^+$ and $B^-$ candidates in the mass range $[5050, 5500]\ \text{MeV}/c^2$, for each decay mode and each year. For both decay modes, the signal shape is described by a Hypatia function [45]; the combinatorial background is modeled by an exponential function; partially reconstructed $B$-meson decays, such as $B \to J/\psi h\pi$ with the $\pi$ meson missing, which contribute in the low-mass region, are described by an ARGUS function [46] convolved with a

TABLE I. Signal yields and raw-charge asymmetries for $B^+ \to J/\psi\pi^+$ and $B^+ \to J/\psi K^+$ decays determined from the mass fits, where the uncertainties are statistical only.

| | 2016 | 2017 | 2018 |
|---|---|---|---|
| $N_{J/\psi\pi}$ | $15\,500 \pm 140$ | $15\,140 \pm 140$ | $18\,130 \pm 150$ |
| $N_{J/\psi K}$ | $371\,700 \pm 600$ | $367\,300 \pm 600$ | $454\,100 \pm 700$ |
| $a^{\rm raw}_{J/\psi\pi}$ $[10^{-2}]$ | $0.91 \pm 0.85$ | $0.50 \pm 0.85$ | $1.42 \pm 0.78$ |
| $a^{\rm raw}_{J/\psi K}$ $[10^{-2}]$ | $-1.35 \pm 0.17$ | $-1.12 \pm 0.17$ | $-1.07 \pm 0.15$ |

Gaussian function. For the CKM-suppressed $B^+ \to J/\psi\pi^+$ mode, a cross-feed background from the favored $B^+ \to J/\psi K^+$ decays with the kaon misidentified as a pion is described by a double-sided crystal ball (DSCB) [47] function. The cross-feed background from $B^+ \to J/\psi\pi^+$ decays is conversely negligible for the $B^+ \to J/\psi K^+$ mass fit. All shape and position parameters are shared between the $B^+$ and $B^-$ decays in the baseline fit. The tail parameters of the Hypatia and DSCB functions are fixed to the values obtained from simulation.

Denoting the signal yields for $B^\pm \to J/\psi h^\pm$ decays as $N_{J/\psi h^\pm}$, their sum $N_{J/\psi h}$, and raw-charge asymmetries $a^{\rm raw}_{J/\psi h} \equiv (N_{J/\psi h^-} - N_{J/\psi h^+})/(N_{J/\psi h^-} + N_{J/\psi h^+})$ are obtained from the mass fits and reported in Table I. Figure 1 shows the mass distributions of the selected $B^\pm \to J/\psi\pi^\pm$ and $B^\pm \to J/\psi K^\pm$ candidates together with the fit projections.

The ratio of the branching fractions of $B^+ \to J/\psi\pi^+$ and $B^+ \to J/\psi K^+$ decays is determined according to

$$\mathcal{R}_{\pi/K} = \frac{N_{J/\psi\pi}}{N_{J/\psi K}} \times \frac{\varepsilon_{J/\psi K}}{\varepsilon_{J/\psi\pi}}, \tag{3}$$

where $\varepsilon_{J/\psi K}$ and $\varepsilon_{J/\psi\pi}$ stand for the total efficiencies, including those of the detector acceptance, trigger, and offline selection. All efficiencies are estimated from simulation after corrections are applied except for the PID efficiency. The latter is obtained for the accompanying hadron using dedicated control samples where pions and kaons are selected without PID requirements and weighted to match the hadron kinematic spectra and event multiplicity in the calibrated signal simulation. The ratio of the total efficiencies, $\varepsilon_{J/\psi\pi}/\varepsilon_{J/\psi K}$, is found to be $0.935 \pm 0.004$, $0.936 \pm 0.004$, and $0.953 \pm 0.005$ for the 2016, 2017, and 2018 data samples, respectively. Here the uncertainties are due to the limited sizes of the simulation and control samples and are propagated to the statistical uncertainties of the $\mathcal{R}_{\pi/K}$ measurements.

The difference in $CP$ asymmetries between $B^+ \to J/\psi\pi^+$ and $B^+ \to J/\psi K^+$ decays is obtained from the raw-charge asymmetries after correcting for the accompanying-hadron detection asymmetries, $a^{\rm det}_{J/\psi h}$, and PID efficiency asymmetries, $a^{\rm PID}_{J/\psi h}$, following

$$\Delta\mathcal{A}^{CP} = \left(a^{\rm raw}_{J/\psi\pi} - a^{\rm raw}_{J/\psi K}\right) - \left(a^{\rm det}_{J/\psi\pi} - a^{\rm det}_{J/\psi K}\right) - \left(a^{\rm PID}_{J/\psi\pi} - a^{\rm PID}_{J/\psi K}\right). \tag{4}$$

The effects of different production cross sections of $B^-$ and $B^+$ mesons are largely canceled in the $\Delta\mathcal{A}^{CP}$ measurement, and further reduced by weighting the $B^+ \to J/\psi K^+$ sample to eliminate a small difference with the $B^+ \to J/\psi\pi^+$ sample in the $B^+$ kinematic distributions. The difference between the pion and kaon detection asymmetries as a function of the kaon momentum is determined from the raw asymmetries of the decays $D^+ \to K^-\pi^+\pi^+$ and $D^+ \to K^0_S\pi^+$ measured with Run 2 data, following the method described in Refs. [48,49]. Using the kaon momentum spectrum in the selected $B^+ \to J/\psi K^+$ sample, the average detection asymmetry difference, common to all years of data taking, is computed to be

$$a^{\rm det}_{J/\psi\pi} - a^{\rm det}_{J/\psi K} = (0.84 \pm 0.05) \times 10^{-2}, \qquad (5)$$

where the uncertainty also accounts for the difference between the pion momentum spectra in $B^+ \to J/\psi\pi^+$ and $D^+ \to K^-\pi^+\pi^+$ decays.

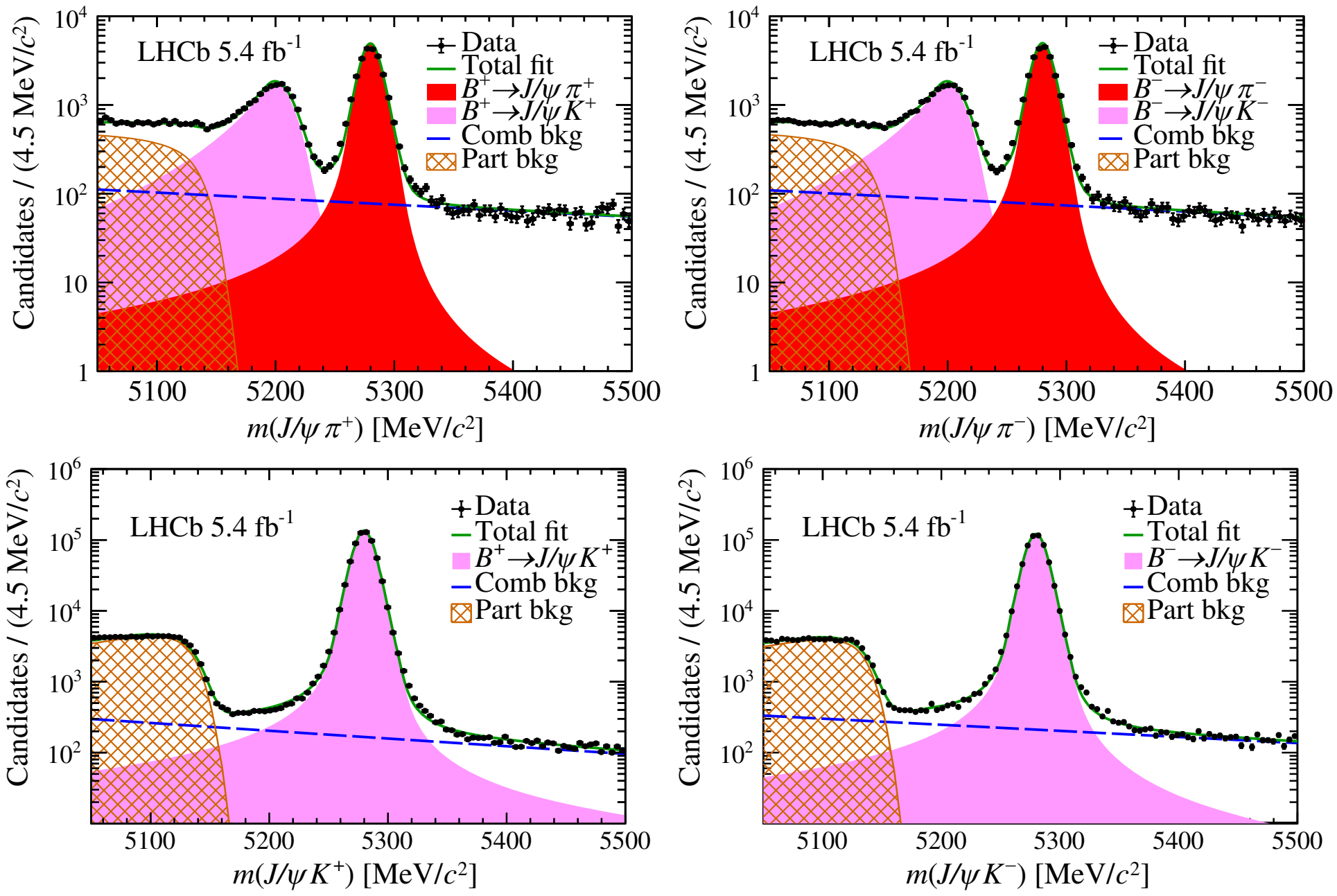

FIG. 1. Mass distributions of the (top left panel) $B^+ \to J/\psi\pi^+$, (top right panel) $B^- \to J/\psi\pi^-$, (bottom left panel) $B^+ \to J/\psi K^+$, and (bottom right panel) $B^- \to J/\psi K^-$ candidates in the combined data sample collected in 2016–2018, with the fit projections also shown.

The hadron PID asymmetries are obtained by separately measuring the PID efficiencies for negative and positive hadrons using Run 2 control samples following the method described in Refs. [38,39]. Their values are

$$a^{\rm PID}_{J/\psi\pi} = \begin{cases} (-0.01 \pm 0.02) \times 10^{-2} & (2016), \\ (+0.00 \pm 0.05) \times 10^{-2} & (2017), \\ (+0.02 \pm 0.06) \times 10^{-2} & (2018), \end{cases} \qquad (6)$$

and

$$a^{\rm PID}_{J/\psi K} = \begin{cases} (+0.00 \pm 0.06) \times 10^{-2} & (2016), \\ (+0.03 \pm 0.05) \times 10^{-2} & (2017), \\ (-0.05 \pm 0.06) \times 10^{-2} & (2018). \end{cases} \qquad (7)$$

The systematic uncertainties in the branching fraction ratio $\mathcal{R}_{\pi/K}$ and $CP$ asymmetry difference $\Delta\mathcal{A}^{CP}$ for each data-taking year are summarized in Table II. Sources of systematic uncertainties associated with the mass fits, the efficiency evaluation, and the nuisance asymmetries are considered. Owing to the inevitability of mass mismodeling whenever such sizable yields are present, associated systematic uncertainties are evaluated by increasing signal and background model sophistication. Mitigating changes include the use of alternative functions to describe the signal and background shapes and fit configurations that allow the position and width parameters of the $B^+$ and $B^-$ signal decays to take separate values within the nominal model. The systematic uncertainty associated with the trigger efficiency is determined by comparing the trigger efficiency ratio between the $B^+ \to J/\psi\pi^+$ and $B^+ \to J/\psi K^+$ modes obtained from simulation to that obtained from control samples consisting of events triggered independently of the signal decays using a data-driven method [50]. The systematic uncertainty due to an imperfect description of the detector material, which affects the $K/\pi$ tracking efficiency from simulation, is estimated by varying the amount of material in the relevant detector volumes by about $\pm 10\%$ [19]. The systematic uncertainty associated with the corrections to the signal simulation is estimated by resampling the relevant simulation and data samples with replacement [51] and repeating the kinematic weighting and efficiency estimation procedure multiple times. The standard deviation of the efficiency ratio $\varepsilon_{J/\psi\pi}/\varepsilon_{J/\psi K}$ is propagated to the $\mathcal{R}_{\pi/K}$ measurement. A systematic uncertainty in $\mathcal{R}_{\pi/K}$ related to the PID efficiency calibration is also evaluated by changing the hadron $p_{\rm T}$ and $\eta$ bin widths used to divide the control samples.

The uncertainties of the estimated detection asymmetry difference in Eq. (5) and PID asymmetries in Eqs. (6) and (7) are propagated to $\Delta\mathcal{A}^{CP}$ as systematic uncertainties. For the baseline result of $\Delta\mathcal{A}^{CP}$, the $B^+ \to J/\psi\pi^+$ sample is weighted to match the kinematic distribution of the $B^+ \to J/\psi K^+$ sample in order to cancel the effect of the $B^+/B^-$ production asymmetry on the measurement. The difference of the $\Delta\mathcal{A}^{CP}$ values obtained with and without this weighting step is taken as a systematic uncertainty.

Robustness of the fit procedure is tested by splitting the data samples according to magnet polarity and by tightening the BDT-output requirements. The results are consistent in all checks.

TABLE II. Relative systematic uncertainties on the branching fraction ratio $\mathcal{R}_{\pi/K}$ and absolute systematic uncertainties on the $CP$-asymmetry difference $\Delta\mathcal{A}^{CP}$ from each source and their quadratic sum.

| | Branching fraction ratio | | | $CP$-asymmetry difference | | |
|---|---|---|---|---|---|---|
| | 2016 (%) | 2017 (%) | 2018 (%) | 2016 ($10^{-2}$) | 2017 ($10^{-2}$) | 2018 ($10^{-2}$) |
| Mass fit | 0.22 | 0.16 | 0.21 | 0.04 | 0.06 | 0.04 |
| Trigger efficiency | 0.40 | 0.39 | 0.37 | ... | ... | ... |
| Material budget | 0.30 | 0.30 | 0.30 | ... | ... | ... |
| Simulation correction | 0.17 | 0.15 | 0.14 | ... | ... | ... |
| PID | 0.29 | 0.22 | 0.29 | 0.06 | 0.07 | 0.08 |
| Detection asymmetry | ... | ... | ... | 0.05 | 0.05 | 0.05 |
| Production asymmetry | ... | ... | ... | 0.02 | 0.02 | 0.02 |
| Total | 0.64 | 0.58 | 0.61 | 0.09 | 0.11 | 0.11 |

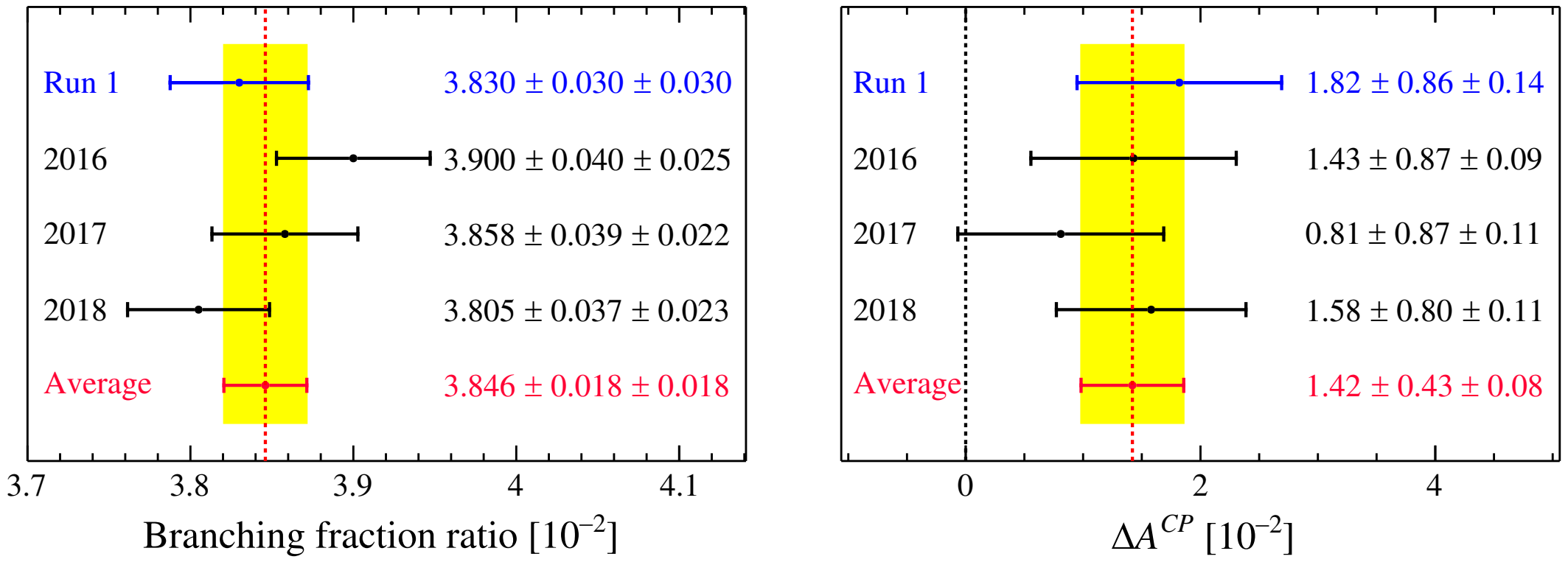

FIG. 2. Comparison of the $\mathcal{R}_{\pi/K}$ and $\Delta\mathcal{A}^{CP}$ measurements from Run 1 2016, 2017, and 2018 data and the average values. The error bars correspond to the sum of the statistical and systematic uncertainties in quadrature.

Using the estimated signal yields, the efficiency ratios, raw-charge and efficiency asymmetries, the ratio of branching fractions, and the $CP$ asymmetry difference between the $B^+ \to J/\psi\pi^+$ and $B^+ \to J/\psi K^+$ decays are determined for each year to be

$$\mathcal{R}_{\pi/K} = \begin{cases} (3.900 \pm 0.040 \pm 0.025) \times 10^{-2} & (2016), \\ (3.858 \pm 0.039 \pm 0.022) \times 10^{-2} & (2017), \\ (3.805 \pm 0.037 \pm 0.023) \times 10^{-2} & (2017), \end{cases}$$

$$\Delta\mathcal{A}^{CP} = \begin{cases} (1.43 \pm 0.87 \pm 0.09) \times 10^{-2} & (2016), \\ (0.81 \pm 0.87 \pm 0.11) \times 10^{-2} & (2017), \\ (1.58 \pm 0.80 \pm 0.11) \times 10^{-2} & (2018), \end{cases}$$

where the first uncertainties are statistical and the second are systematic. The measurements for different years are consistent with each other and combined using the best linear unbiased estimator method [52,53] to obtain

$$\mathcal{R}_{\pi/K} = (3.852 \pm 0.022 \pm 0.018) \times 10^{-2},$$
$$\Delta\mathcal{A}^{CP} = (1.29 \pm 0.49 \pm 0.08) \times 10^{-2}.$$

The Run 2 results are further combined with the LHCb Run 1 measurements [19] using the same method, yielding

$$\mathcal{R}_{\pi/K} = (3.846 \pm 0.018 \pm 0.018) \times 10^{-2},$$
$$\Delta\mathcal{A}^{CP} = (1.42 \pm 0.43 \pm 0.08) \times 10^{-2}.$$

In the above combinations, the systematic uncertainties related to the material budget and hadron detection asymmetries are considered to be fully correlated between different data-taking periods. The inputs and outcomes of the combination are displayed in Fig. 2. The significance of the nonzero $\Delta\mathcal{A}^{CP}$ value is evaluated to be 3.2 standard deviations ($\sigma$), representing the first evidence for direct $CP$ violation in beauty to charmonium decays.

Using the LHCb measurement of the $CP$ asymmetry in the $B^+ \to J/\psi K^+$ decay [49] and taking into account its correlation with the $\Delta\mathcal{A}^{CP}$ measurement from this analysis due to the overlap in datasets, the $CP$ asymmetry in the $B^+ \to J/\psi\pi^+$ decay is determined to be $\mathcal{A}^{CP}(B^+ \to J/\psi\pi^+) = (1.51 \pm 0.50 \pm 0.08) \times 10^{-2}$.

In summary, the most precise measurements of the $CP$-asymmetry difference between $B^+ \to J/\psi\pi^+$ and $B^+ \to J/\psi K^+$ decays and their branching fraction ratio are obtained using LHCb data collected at 13 TeV in 2016–2018, corresponding to an integrated luminosity of 5.4 $\mathrm{fb}^{-1}$. These results are then combined with the previous LHCb Run 1 measurements. The combined $CP$ asymmetry difference shows a 3.2$\sigma$ deviation from zero, providing the first evidence for direct $CP$ violation in beauty decays to charmonium final states. This effect can be attributed to the enhanced penguin to tree ratio in $B^+ \to J/\psi\pi^+$ decays compared to that in $b \to c\bar{c}s$ transitions. The $\Delta\mathcal{A}^{CP}$ and $\mathcal{R}_{\pi/K}$ measurements serve to control the effects of the penguin contributions in the golden channel $B^0 \to J/\psi K^0$ affecting the determination of the $CP$-violating phase $2\beta$, using approximate SU(3) flavor symmetry. Constraints on the size and strong phase of the penguin contribution relative to the tree obtained using the $\Delta\mathcal{A}^{CP}$ and $\mathcal{R}_{\pi/K}$ measurements can be found in the Supplemental Material [54].

*Acknowledgments*—We express our gratitude to our colleagues in the CERN accelerator departments for the excellent performance of the LHC. We thank the technical and administrative staff at the LHCb institutes. We acknowledge support from CERN and from the national agencies CAPES, CNPq, FAPERJ, and FINEP (Brazil); MOST and NSFC (China); CNRS/IN2P3 (France); BMBF, DFG, and MPG (Germany); INFN (Italy); NWO (Netherlands); MNiSW and NCN (Poland); MCID/IFA (Romania); MICIU and AEI (Spain); SNSF and SER (Switzerland); NASU (Ukraine); STFC (United Kingdom); and DOE NP and NSF (USA). We acknowledge the computing resources that are provided by CERN, IN2P3 (France), KIT and DESY (Germany), INFN (Italy), SURF (Netherlands), PIC (Spain), GridPP (United Kingdom), CSCS (Switzerland), IFIN-HH (Romania), CBPF (Brazil), and Polish WLCG (Poland). We are indebted to the communities behind the multiple open-source software packages on which we depend. Individual groups or members have received support from ARC and ARDC (Australia); the Key Research Program of Frontier Sciences of CAS, CAS PIFI, CAS CCEPP, Fundamental Research Funds for the Central Universities, and Science and Technology Program of Guangzhou (China); Minciencias (Colombia); EPLANET, Marie Skłodowska-Curie Actions, ERC, and NextGenerationEU (European Union); A*MIDEX, ANR, IPhU, and Labex P2IO, and Région Auvergne-Rhône-Alpes (France); AvH Foundation (Germany); ICSC (Italy); Severo Ochoa and María de Maeztu Units of Excellence, GVA, XuntaGal, GENCAT, InTalent-Inditex and Programa Atracción Talento CM (Spain); SRC (Sweden); the Leverhulme Trust, the Royal Society, and UKRI (United Kingdom).

*Data availability*—The data that support the findings of this article are openly available [55].

R. Aaij,[38] A. S. W. Abdelmotteleb,[57] C. Abellan Beteta,[51] F. Abudinén,[57] T. Ackernley,[61] A. A. Adefisoye,[69] B. Adeva,[47] M. Adinolfi,[55] P. Adlarson,[82] C. Agapopoulou,[14] C. A. Aidala,[83] Z. Ajaltouni,[11] S. Akar,[66] K. Akiba,[38] P. Albicocco,[28] J. Albrecht,[19,b] F. Alessio,[49] M. Alexander,[60] Z. Aliouche,[63] P. Alvarez Cartelle,[56] R. Amalric,[16] S. Amato,[3] J. L. Amey,[55] Y. Amhis,[14,49] L. An,[6] L. Anderlini,[27] M. Andersson,[51] A. Andreianov,[44] P. Andreola,[51] M. Andreotti,[26] D. Andreou,[69] A. Anelli,[31,c] D. Ao,[7] F. Archilli,[37,d] M. Argenton,[26] S. Arguedas Cuendis,[9,49] A. Artamonov,[44] M. Artuso,[69] E. Aslanides,[13] R. Ataíde Da Silva,[50] M. Atzeni,[65] B. Audurier,[12] D. Bacher,[64] I. Bachiller Perea,[10] S. Bachmann,[22] M. Bachmayer,[50] J. J. Back,[57] P. Baladron Rodriguez,[47] V. Balagura,[15] A. Balboni,[26] W. Baldini,[26] L. Balzani,[19] H. Bao,[7] J. Baptista de Souza Leite,[61] C. Barbero Pretel,[47,12] M. Barbetti,[27] I. R. Barbosa,[70] R. J. Barlow,[63] M. Barnyakov,[25] S. Barsuk,[14] W. Barter,[59] M. Bartolini,[56] J. Bartz,[69] J. M. Basels,[17] S. Bashir,[40] G. Bassi,[35,e] B. Batsukh,[5] P. B. Battista,[14] A. Bay,[50] A. Beck,[57] M. Becker,[19] F. Bedeschi,[35] I. B. Bediaga,[2] N. A. Behling,[19] S. Belin,[47] V. Bellee,[51] K. Belous,[44] I. Belov,[29] I. Belyaev,[36] G. Benane,[13] G. Bencivenni,[28] E. Ben-Haim,[16] A. Berezhnoy,[44] R. Bernet,[51] S. Bernet Andres,[45] A. Bertolin,[33] C. Betancourt,[51] F. Betti,[59] J. Bex,[56] Ia. Bezshyiko,[51] J. Bhom,[41] M. S. Bieker,[19] N. V. Biesuz,[26] P. Billoir,[16] A. Biolchini,[38] M. Birch,[62] F. C. R. Bishop,[10] A. Bitadze,[63] A. Bizzeti, T. Blake,[57] F. Blanc,[50] J. E. Blank,[19] S. Blusk,[69] V. Bocharnikov,[44]

J. A. Boelhauve,[19] O. Boente Garcia,[15] T. Boettcher,[66] A. Bohare,[59] A. Boldyrev,[44] C. S. Bolognani,[79] R. Bolzonella,[26,f] R. B. Bonacci,[1] N. Bondar,[44] A. Bordelius,[49] F. Borgato,[33,g] S. Borghi,[63] M. Borsato,[31,c] J. T. Borsuk,[41] S. A. Bouchiba,[50] M. Bovill,[64] T. J. V. Bowcock,[61] A. Boyer,[49] C. Bozzi,[26] A. Brea Rodriguez,[50] N. Breer,[19] J. Brodzicka,[41] A. Brossa Gonzalo,[47,a] J. Brown,[61] D. Brundu,[32] E. Buchanan,[59] A. Buonaura,[51] L. Buonincontri,[33,g] A. T. Burke,[63] C. Burr,[49] J. S. Butter,[56] J. Buytaert,[49] W. Byczynski,[49] S. Cadeddu,[32] H. Cai,[74] A. C. Caillet,[16] R. Calabrese,[26,f] S. Calderon Ramirez,[9] L. Calefice,[46] S. Cali,[28] M. Calvi,[31,c] M. Calvo Gomez,[45] P. Camargo Magalhaes,[2,h] J. I. Cambon Bouzas,[47] P. Campana,[28] D. H. Campora Perez,[79] A. F. Campoverde Quezada,[7] S. Capelli,[31] L. Capriotti,[26] R. Caravaca-Mora,[9] A. Carbone,[25,i] L. Carcedo Salgado,[47] R. Cardinale,[29,j] A. Cardini,[32] P. Carniti,[31,c] L. Carus,[22] A. Casais Vidal,[65] R. Caspary,[22] G. Casse,[61] M. Cattaneo,[49] G. Cavallero,[26,49] V. Cavallini,[26,f] S. Celani,[22] D. Cervenkov,[64] S. Cesare,[30,k] A. J. Chadwick,[61] I. Chahrour,[83] M. Charles,[16] Ph. Charpentier,[49] E. Chatzianagnostou,[38] M. Chefdeville,[10] C. Chen,[13] S. Chen,[5] Z. Chen,[7] A. Chernov,[41] S. Chernyshenko,[53] X. Chiotopoulos,[79] V. Chobanova,[81] S. Cholak,[50] M. Chrzaszcz,[41] A. Chubykin,[44] V. Chulikov,[28] P. Ciambrone,[28] X. Cid Vidal,[47] G. Ciezarek,[49] P. Cifra,[49] P. E. L. Clarke,[59] M. Clemencic,[49] H. V. Cliff,[56] J. Closier,[49] C. Cocha Toapaxi,[22] V. Coco,[49] J. Cogan,[13] E. Cogneras,[11] L. Cojocariu,[43] P. Collins,[49] T. Colombo,[49] M. Colonna,[19] A. Comerma-Montells,[46] L. Congedo,[24] A. Contu,[32] N. Cooke,[60] I. Corredoira,[47] A. Correia,[16] G. Corti,[49] J. J. Cottee Meldrum,[55] B. Couturier,[49] D. C. Craik,[51] M. Cruz Torres,[2,l] E. Curras Rivera,[50] R. Currie,[59] C. L. Da Silva,[68] S. Dadabaev,[44] L. Dai,[71] X. Dai,[6] E. Dall'Occo,[49] J. Dalseno,[47] C. D'Ambrosio,[49] J. Daniel,[11] A. Danilina,[44] P. d'Argent,[24] A. Davidson,[57] J. E. Davies,[63] A. Davis,[63] O. De Aguiar Francisco,[63] C. De Angelis,[32,m] F. De Benedetti,[49] J. de Boer,[38] K. De Bruyn,[78] S. De Capua,[63] M. De Cian,[22,49] U. De Freitas Carneiro Da Graca,[2,n] E. De Lucia,[28] J. M. De Miranda,[2] L. De Paula,[3] M. De Serio,[24,o] P. De Simone,[28] F. De Vellis,[19] J. A. de Vries,[79] F. Debernardis,[24] D. Decamp,[10] V. Dedu,[13] S. Dekkers,[1] L. Del Buono,[16] B. Delaney,[65] H.-P. Dembinski,[19] J. Deng,[8] V. Denysenko,[51] O. Deschamps,[11] F. Dettori,[32,m] B. Dey,[77] P. Di Nezza,[28] I. Diachkov,[44] S. Didenko,[44] S. Ding,[69] L. Dittmann,[22] V. Dobishuk,[53] A. D. Docheva,[60] C. Dong,[4,p] A. M. Donohoe,[23] F. Dordei,[32] A. C. dos Reis,[2] A. D. Dowling,[69] W. Duan,[72] P. Duda,[80] M. W. Dudek,[41] L. Dufour,[49] V. Duk,[34] P. Durante,[49] M. M. Duras,[80] J. M. Durham,[68] O. D. Durmus,[77] A. Dziurda,[41] A. Dzyuba,[44] S. Easo,[58] E. Eckstein,[18] U. Egede,[1] A. Egorychev,[44] V. Egorychev,[44] S. Eisenhardt,[59] E. Ejopu,[63] L. Eklund,[82] M. Elashri,[66] J. Ellbracht,[19] S. Ely,[62] A. Ene,[43] J. Eschle,[69] S. Esen,[22] T. Evans,[63] F. Fabiano,[32,m] L. N. Falcao,[2] Y. Fan,[7] B. Fang,[7] L. Fantini,[34,49,q] M. Faria,[50] K. Farmer,[59] D. Fazzini,[31,c] L. Felkowski,[80] M. Feng,[5,7] M. Feo,[19,49] A. Fernandez Casani,[48] M. Fernandez Gomez,[47] A. D. Fernez,[67] F. Ferrari,[25] F. Ferreira Rodrigues,[3] M. Ferrillo,[51] M. Ferro-Luzzi,[49] S. Filippov,[44] R. A. Fini,[24] M. Fiorini,[26,f] M. Firlej,[40] K. L. Fischer,[64] D. S. Fitzgerald,[83] C. Fitzpatrick,[63] T. Fiutowski,[40] F. Fleuret,[15] M. Fontana,[25] L. F. Foreman,[63] R. Forty,[49] D. Foulds-Holt,[56] V. Franco Lima,[3] M. Franco Sevilla,[67] M. Frank,[49] E. Franzoso,[26,f] G. Frau,[63] C. Frei,[49] D. A. Friday,[63] J. Fu,[7] Q. Führing,[19,56,b] Y. Fujii,[1] T. Fulghesu,[16] E. Gabriel,[38] G. Galati,[24] M. D. Galati,[38] A. Gallas Torreira,[47] D. Galli,[25,i] S. Gambetta,[59] M. Gandelman,[3] P. Gandini,[30] B. Ganie,[63] H. Gao,[7] R. Gao,[64] T. Q. Gao,[56] Y. Gao,[8] Y. Gao,[6] Y. Gao,[8] L. M. Garcia Martin,[50] P. Garcia Moreno,[46] J. García Pardiñas,[49] K. G. Garg,[8] L. Garrido,[46] C. Gaspar,[49] R. E. Geertsema,[38] L. L. Gerken,[19] E. Gersabeck,[63] M. Gersabeck,[20] T. Gershon,[57] S. Ghizzo,[29,j] Z. Ghorbanimoghaddam,[55] L. Giambastiani,[33,g] F. I. Giasemis,[16,r] V. Gibson,[56] H. K. Giemza,[42] A. L. Gilman,[64] M. Giovannetti,[28] A. Gioventù,[46] L. Girardey,[63] P. Gironella Gironell,[46] C. Giugliano,[26,f] M. A. Giza,[41] E. L. Gkougkousis,[62] F. C. Glaser,[14,22] V. V. Gligorov,[16,49] C. Göbel,[70] E. Golobardes,[45] D. Golubkov,[44] A. Golutvin,[62,44,49] S. Gomez Fernandez,[46] W. Gomulka,[40] F. Goncalves Abrantes,[64] M. Goncerz,[41] G. Gong,[4,p] J. A. Gooding,[19] I. V. Gorelov,[44] C. Gotti,[31] J. P. Grabowski,[18] L. A. Granado Cardoso,[49] E. Graugés,[46] E. Graverini,[50,s] L. Grazette,[57] G. Graziani, A. T. Grecu,[43] L. M. Greeven,[38] N. A. Grieser,[66] L. Grillo,[60] S. Gromov,[44] C. Gu,[15] M. Guarise,[26] L. Guerry,[11] M. Guittiere,[14] V. Guliaeva,[44] P. A. Günther,[22] A.-K. Guseinov,[50] E. Gushchin,[44] Y. Guz,[6,49,44] T. Gys,[49] K. Habermann,[18] T. Hadavizadeh,[1] C. Hadjivasiliou,[67] G. Haefeli,[50] C. Haen,[49] J. Haimberger,[49] M. Hajheidari,[49] G. Hallett,[57] M. M. Halvorsen,[49] P. M. Hamilton,[67] J. Hammerich,[61] Q. Han,[8] X. Han,[22,49] S. Hansmann-Menzemer,[22] L. Hao,[7] N. Harnew,[64] M. Hartmann,[14] S. Hashmi,[40] J. He,[7,t] F. Hemmer,[49] C. Henderson,[66] R. D. L. Henderson,[1,57]

A. M. Hennequin,[49] K. Hennessy,[61] L. Henry,[50] J. Herd,[62] P. Herrero Gascon,[22] J. Heuel,[17] A. Hicheur,[3] G. Hijano Mendizabal,[51] D. Hill,[50] J. Horswill,[63] R. Hou,[8] Y. Hou,[11] N. Howarth,[61] J. Hu,[72] W. Hu,[6] X. Hu,[4,p] W. Huang,[7] W. Hulsbergen,[38] R. J. Hunter,[57] M. Hushchyn,[44] D. Hutchcroft,[61] M. Idzik,[40] D. Ilin,[44] P. Ilten,[66] A. Inglessi,[44] A. Iniukhin,[44] A. Ishteev,[44] K. Ivshin,[44] R. Jacobsson,[49] H. Jage,[17] S. J. Jaimes Elles,[48,75] S. Jakobsen,[49] E. Jans,[38] B. K. Jashal,[48] A. Jawahery,[67,49] V. Jevtic,[19] E. Jiang,[67] X. Jiang,[5,7] Y. Jiang,[7] Y. J. Jiang,[6] M. John,[64] A. John Rubesh Rajan,[23] D. Johnson,[54] C. R. Jones,[56] T. P. Jones,[57] S. Joshi,[42] B. Jost,[49] J. Juan Castella,[56] N. Jurik,[49] I. Juszczak,[41] D. Kaminaris,[50] S. Kandybei,[52] M. Kane,[59] Y. Kang,[4,p] C. Kar,[11] M. Karacson,[49] D. Karpenkov,[44] A. Kauniskangas,[50] J. W. Kautz,[66] M. K. Kazanecki,[41] F. Keizer,[49] M. Kenzie,[56] T. Ketel,[38] B. Khanji,[69] A. Kharisova,[44] S. Kholodenko,[35,49] G. Khreich,[14] T. Kirn,[17] V. S. Kirsebom,[31,c] O. Kitouni,[65] S. Klaver,[39] N. Kleijne,[35,e] K. Klimaszewski,[42] M. R. Kmiec,[42] S. Koliiev,[53] L. Kolk,[19] A. Konoplyannikov,[44] P. Kopciewicz,[40,49] P. Koppenburg,[38] M. Korolev,[44] I. Kostiuk,[38] O. Kot,[53] S. Kotriakhova, A. Kozachuk,[44] P. Kravchenko,[44] L. Kravchuk,[44] M. Kreps,[57] P. Krokovny,[44] W. Krupa,[69] W. Krzemien,[42] O. Kshyvanskyi,[53] S. Kubis,[80] M. Kucharczyk,[41] V. Kudryavtsev,[44] E. Kulikova,[44] A. Kupsc,[82] B. K. Kutsenko,[13] D. Lacarrere,[49] P. Laguarta Gonzalez,[46] A. Lai,[32] A. Lampis,[32] D. Lancierini,[56] C. Landesa Gomez,[47] J. J. Lane,[1] R. Lane,[55] G. Lanfranchi,[28] C. Langenbruch,[22] J. Langer,[19] O. Lantwin,[44] T. Latham,[57] F. Lazzari,[35,s] C. Lazzeroni,[54] R. Le Gac,[13] H. Lee,[61] R. Lefèvre,[11] A. Leflat,[44] S. Legotin,[44] M. Lehuraux,[57] E. Lemos Cid,[49] O. Leroy,[13] T. Lesiak,[41] E. D. Lesser,[49] B. Leverington,[22] A. Li,[4,p] C. Li,[13] H. Li,[72] K. Li,[8] L. Li,[63] M. Li,[8] P. Li,[7] P.-R. Li,[73] Q. Li,[5,7] S. Li,[8] T. Li,[5,u] T. Li,[72] Y. Li,[8] Y. Li,[5] Z. Lian,[4,p] X. Liang,[69] S. Libralon,[48] C. Lin,[7] T. Lin,[58] R. Lindner,[49] H. Linton,[62] V. Lisovskyi,[50] R. Litvinov,[32,49] F. L. Liu,[1] G. Liu,[72] K. Liu,[73] S. Liu,[5,7] W. Liu,[8] Y. Liu,[59] Y. Liu,[73] Y. L. Liu,[62] A. Lobo Salvia,[46] A. Loi,[32] J. Lomba Castro,[47] T. Long,[56] J. H. Lopes,[3] A. Lopez Huertas,[46] S. López Soliño,[47] Q. Lu,[15] C. Lucarelli,[27] D. Lucchesi,[33,g] M. Lucio Martinez,[79] V. Lukashenko,[38,53] Y. Luo,[6] A. Lupato,[33,v] E. Luppi,[26,f] K. Lynch,[23] X.-R. Lyu,[7] G. M. Ma,[4,p] S. Maccolini,[19] F. Machefert,[14] F. Maciuc,[43] B. Mack,[69] I. Mackay,[64] L. M. Mackey,[69] L. R. Madhan Mohan,[56] M. J. Madurai,[54] A. Maevskiy,[44] D. Magdalinski,[38] D. Maisuzenko,[44] M. W. Majewski,[40] J. J. Malczewski,[41] S. Malde,[64] L. Malentacca,[49] A. Malinin,[44] T. Maltsev,[44] G. Manca,[32,m] G. Mancinelli,[13] C. Mancuso,[30,14,k] R. Manera Escalero,[46] F. M. Manganella,[37] D. Manuzzi,[25] D. Marangotto,[30,k] J. F. Marchand,[10] R. Marchevski,[50] U. Marconi,[25] E. Mariani,[16] S. Mariani,[49] C. Marin Benito,[46,49] J. Marks,[22] A. M. Marshall,[55] L. Martel,[64] G. Martelli,[34,q] G. Martellotti,[36] L. Martinazzoli,[49] M. Martinelli,[31,c] D. Martinez Gomez,[78] D. Martinez Santos,[81] F. Martinez Vidal,[48] A. Martorell i Granollers,[45] A. Massafferri,[2] R. Matev,[49] A. Mathad,[49] V. Matiunin,[44] C. Matteuzzi,[69] K. R. Mattioli,[15] A. Mauri,[62] E. Maurice,[15] J. Mauricio,[46] P. Mayencourt,[50] J. Mazorra de Cos,[48] M. Mazurek,[42] M. McCann,[62] L. Mcconnell,[23] T. H. McGrath,[63] N. T. McHugh,[60] A. McNab,[63] R. McNulty,[23] B. Meadows,[66] G. Meier,[19] D. Melnychuk,[42] F. M. Meng,[4,p] M. Merk,[38,79] A. Merli,[50] L. Meyer Garcia,[67] D. Miao,[5,7] H. Miao,[7] M. Mikhasenko,[76] D. A. Milanes,[75] A. Minotti,[31,c] E. Minucci,[28] T. Miralles,[11] B. Mitreska,[19] D. S. Mitzel,[19] A. Modak,[58] R. A. Mohammed,[64] R. D. Moise,[17] S. Mokhnenko,[44] E. F. Molina Cardenas,[83] T. Mombächer,[49] M. Monk,[57,1] S. Monteil,[11] A. Morcillo Gomez,[47] G. Morello,[28] M. J. Morello,[35,e] M. P. Morgenthaler,[22] J. Moron,[40] W. Morren,[38] A. B. Morris,[49] A. G. Morris,[13] R. Mountain,[69] H. Mu,[4,p] Z. M. Mu,[6] E. Muhammad,[57] F. Muheim,[59] M. Mulder,[78] K. Müller,[51] F. Muñoz-Rojas,[9] R. Murta,[62] P. Naik,[61] T. Nakada,[50] R. Nandakumar,[58] T. Nanut,[49] I. Nasteva,[3] M. Needham,[59] N. Neri,[30,k] S. Neubert,[18] N. Neufeld,[49] P. Neustroev,[44] J. Nicolini,[19,14] D. Nicotra,[79] E. M. Niel,[50] N. Nikitin,[44] Q. Niu,[73] P. Nogarolli,[3] P. Nogga,[18] C. Normand,[55] J. Novoa Fernandez,[47] G. Nowak,[66] C. Nunez,[83] H. N. Nur,[60] A. Oblakowska-Mucha,[40] V. Obraztsov,[44] T. Oeser,[17] S. Okamura,[26,f] A. Okhotnikov,[44] O. Okhrimenko,[53] R. Oldeman,[32,m] F. Oliva,[59] M. Olocco,[19] C. J. G. Onderwater,[79] R. H. O'Neil,[59] D. Osthues,[19] J. M. Otalora Goicochea,[3] P. Owen,[51] A. Oyanguren,[48] O. Ozcelik,[59] F. Paciolla,[35,w] A. Padee,[42] K. O. Padeken,[18] B. Pagare,[57] P. R. Pais,[22] T. Pajero,[49] A. Palano,[24] M. Palutan,[28] X. Pan,[4,p] G. Panshin,[44] L. Paolucci,[57] A. Papanestis,[58,49] M. Pappagallo,[24,o] L. L. Pappalardo,[26,f] C. Pappenheimer,[66] C. Parkes,[63] B. Passalacqua,[26,f] G. Passaleva,[27] D. Passaro,[35,e] A. Pastore,[24] M. Patel,[62] J. Patoc,[64] C. Patrignani,[25,i] A. Paul,[69] C. J. Pawley,[79] A. Pellegrino,[38] J. Peng,[5,7] M. Pepe Altarelli,[28] S. Perazzini,[25] D. Pereima,[44] H. Pereira Da Costa,[68] A. Pereiro Castro,[47] P. Perret,[11] A. Perrevoort,[78]

A. Perro,[49,13] K. Petridis,[55] A. Petrolini,[29,j] J. P. Pfaller,[66] H. Pham,[69] L. Pica,[35,e] M. Piccini,[34] L. Piccolo,[32] B. Pietrzyk,[10] G. Pietrzyk,[14] D. Pinci,[36] F. Pisani,[49] M. Pizzichemi,[31,49,c] V. Placinta,[43] M. Plo Casasus,[47] T. Poeschl,[49] F. Polci,[16,49] M. Poli Lener,[28] A. Poluektov,[13] N. Polukhina,[44] I. Polyakov,[44] E. Polycarpo,[3] S. Ponce,[49] D. Popov,[7] S. Poslavskii,[44] K. Prasanth,[59] C. Prouve,[47] D. Provenzano,[32,m] V. Pugatch,[53] G. Punzi,[35,s] S. Qasim,[51] Q. Q. Qian,[6] W. Qian,[7] N. Qin,[4,p] S. Qu,[4,p] R. Quagliani,[49] R. I. Rabadan Trejo,[57] J. H. Rademacker,[55] M. Rama,[35] M. Ramírez García,[83] V. Ramos De Oliveira,[70] M. Ramos Pernas,[57] M. S. Rangel,[3] F. Ratnikov,[44] G. Raven,[39] M. Rebollo De Miguel,[48] F. Redi,[30,v] J. Reich,[55] F. Reiss,[63] Z. Ren,[7] P. K. Resmi,[64] R. Ribatti,[50] G. R. Ricart,[15,12] D. Riccardi,[35,e] S. Ricciardi,[58] K. Richardson,[65] M. Richardson-Slipper,[59] K. Rinnert,[61] P. Robbe,[14,49] G. Robertson,[60] E. Rodrigues,[61] A. Rodriguez Alvarez,[46] E. Rodriguez Fernandez,[47] J. A. Rodriguez Lopez,[75] E. Rodriguez Rodriguez,[47] J. Roensch,[19] A. Rogachev,[44] A. Rogovskiy,[58] D. L. Rolf,[49] P. Roloff,[49] V. Romanovskiy,[66] A. Romero Vidal,[47] G. Romolini,[26] F. Ronchetti,[50] T. Rong,[6] M. Rotondo,[28] S. R. Roy,[22] M. S. Rudolph,[69] M. Ruiz Diaz,[22] R. A. Ruiz Fernandez,[47] J. Ruiz Vidal,[82,x] A. Ryzhikov,[44] J. Ryzka,[40] J. J. Saavedra-Arias,[9] J. J. Saborido Silva,[47] R. Sadek,[15] N. Sagidova,[44] D. Sahoo,[77] N. Sahoo,[54] B. Saitta,[32,m] M. Salomoni,[31,49,c] I. Sanderswood,[48] R. Santacesaria,[36] C. Santamarina Rios,[47] M. Santimaria,[28,49] L. Santoro,[2] E. Santovetti,[37] A. Saputi,[26,49] D. Saranin,[44] A. Sarnatskiy,[78] G. Sarpis,[59] M. Sarpis,[63] C. Satriano,[36,y] A. Satta,[37] M. Saur,[6] D. Savrina,[44] H. Sazak,[17] F. Sborzacchi,[49,28] L. G. Scantlebury Smead,[64] A. Scarabotto,[19] S. Schael,[17] S. Scherl,[61] M. Schiller,[60] H. Schindler,[49] M. Schmelling,[21] B. Schmidt,[49] S. Schmitt,[17] H. Schmitz,[18] O. Schneider,[50] A. Schopper,[49] N. Schulte,[19] S. Schulte,[50] M. H. Schune,[14] R. Schwemmer,[49] G. Schwering,[17] B. Sciascia,[28] A. Sciuccati,[49] S. Sellam,[47] A. Semennikov,[44] T. Senger,[51] M. Senghi Soares,[39] A. Sergi,[29,j] N. Serra,[51] L. Sestini,[33] A. Seuthe,[19] Y. Shang,[6] D. M. Shangase,[83] M. Shapkin,[44] R. S. Sharma,[69] I. Shchemerov,[44] L. Shchutska,[50] T. Shears,[61] L. Shekhtman,[44] Z. Shen,[6] S. Sheng,[5,7] V. Shevchenko,[44] B. Shi,[7] Q. Shi,[7] Y. Shimizu,[14] E. Shmanin,[25] R. Shorkin,[44] J. D. Shupperd,[69] R. Silva Coutinho,[69] G. Simi,[33,g] S. Simone,[24,o] N. Skidmore,[57] T. Skwarnicki,[69] M. W. Slater,[54] J. C. Smallwood,[64] E. Smith,[65] K. Smith,[68] M. Smith,[62] A. Snoch,[38] L. Soares Lavra,[59] M. D. Sokoloff,[66] F. J. P. Soler,[60] A. Solomin,[44,55] A. Solovev,[44] I. Solovyev,[44] N. S. Sommerfeld,[18] R. Song,[1] Y. Song,[50] Y. Song,[4,p] Y. S. Song,[6] F. L. Souza De Almeida,[69] B. Souza De Paula,[3] E. Spadaro Norella,[29,j] E. Spedicato,[25] J. G. Speer,[19] E. Spiridenkov,[44] P. Spradlin,[60] V. Sriskaran,[49] F. Stagni,[49] M. Stahl,[49] S. Stahl,[49] S. Stanislaus,[64] E. N. Stein,[49] O. Steinkamp,[51] O. Stenyakin,[44] H. Stevens,[19] D. Strekalina,[44] Y. Su,[7] F. Suljik,[64] J. Sun,[32] L. Sun,[74] D. Sundfeld,[2] W. Sutcliffe,[51] P. N. Swallow,[54] K. Swientek,[40] F. Swystun,[56] A. Szabelski,[42] T. Szumlak,[40] Y. Tan,[4,p] Y. Tang,[74] M. D. Tat,[64] A. Terentev,[44] F. Terzuoli,[35,49,w] F. Teubert,[49] E. Thomas,[49] D. J. D. Thompson,[54] H. Tilquin,[62] V. Tisserand,[11] S. T'Jampens,[10] M. Tobin,[5,49] L. Tomassetti,[26,f] G. Tonani,[30,49,k] X. Tong,[6] D. Torres Machado,[2] L. Toscano,[19] D. Y. Tou,[4,p] C. Trippl,[45] G. Tuci,[22] N. Tuning,[38] L. H. Uecker,[22] A. Ukleja,[40] D. J. Unverzagt,[22] B. Urbach,[59] E. Ursov,[44] A. Usachov,[39] A. Ustyuzhanin,[44] U. Uwer,[22] V. Vagnoni,[25] V. Valcarce Cadenas,[47] G. Valenti,[25] N. Valls Canudas,[49] H. Van Hecke,[68] E. van Herwijnen,[62] C. B. Van Hulse,[47,z] R. Van Laak,[50] M. van Veghel,[38] G. Vasquez,[51] R. Vazquez Gomez,[46] P. Vazquez Regueiro,[47] C. Vázquez Sierra,[47] S. Vecchi,[26] J. J. Velthuis,[55] M. Veltri,[27,aa] A. Venkateswaran,[50] M. Verdoglia,[32] M. Vesterinen,[57] D. Vico Benet,[64] P. Vidrier Villalba,[46] M. Vieites Diaz,[49] X. Vilasis-Cardona,[45] E. Vilella Figueras,[61] A. Villa,[25] P. Vincent,[16] F. C. Volle,[54] D. vom Bruch,[13] N. Voropaev,[44] K. Vos,[79] G. Vouters,[10] C. Vrahas,[59] J. Wagner,[19] J. Walsh,[35] E. J. Walton,[1,57] G. Wan,[6] C. Wang,[22] G. Wang,[8] H. Wang,[73] J. Wang,[6] J. Wang,[5] J. Wang,[4,p] J. Wang,[74] M. Wang,[30] N. W. Wang,[7] R. Wang,[55] X. Wang,[8] X. Wang,[72] X. W. Wang,[62] Y. Wang,[6] Y. W. Wang,[73] Z. Wang,[14] Z. Wang,[4,p] Z. Wang,[30] J. A. Ward,[57,1] M. Waterlaat,[49] N. K. Watson,[54] D. Websdale,[62] Y. Wei,[6] J. Wendel,[81] B. D. C. Westhenry,[55] C. White,[56] M. Whitehead,[60] E. Whiter,[54] A. R. Wiederhold,[63] D. Wiedner,[19] G. Wilkinson,[64] M. K. Wilkinson,[66] M. Williams,[65] M. J. Williams,[49] M. R. J. Williams,[59] R. Williams,[56] Z. Williams,[55] F. F. Wilson,[58] M. Winn,[12] W. Wislicki,[42] M. Witek,[41] L. Witola,[22] G. Wormser,[14] S. A. Wotton,[56] H. Wu,[69] J. Wu,[8] X. Wu,[74] Y. Wu,[6] Z. Wu,[7] K. Wyllie,[49] S. Xian,[72] Z. Xiang,[5] Y. Xie,[8] A. Xu,[35] J. Xu,[7] L. Xu,[4,p] L. Xu,[4,p] M. Xu,[57] Z. Xu,[49] Z. Xu,[7] Z. Xu,[5] D. Yang,[4] K. Yang,[62] S. Yang,[7] X. Yang,[6] Y. Yang,[29,j] Z. Yang,[6] Z. Yang,[67] V. Yeroshenko,[14] H. Yeung,[63] H. Yin,[8] X. Yin,[7] C. Y. Yu,[6] J. Yu,[71] X. Yuan,[5] Y Yuan,[5,7] E. Zaffaroni,[50] M. Zavertyaev,[21] M. Zdybal,[41] F. Zenesini,[25,i] C. Zeng,[5,7]

M. Zeng,[4,p] C. Zhang,[6] D. Zhang,[8] J. Zhang,[7] L. Zhang,[4,p] S. Zhang,[71] S. Zhang,[64] Y. Zhang,[6] Y. Z. Zhang,[4,p] Y. Zhao,[22] A. Zharkova,[44] A. Zhelezov,[22] S. Z. Zheng,[6] X. Z. Zheng,[4,p] Y. Zheng,[7] T. Zhou,[6] X. Zhou,[8] Y. Zhou,[7] V. Zhovkovska,[57] L. Z. Zhu,[7] X. Zhu,[4,p] X. Zhu,[8] V. Zhukov,[17] J. Zhuo,[48] Q. Zou,[5,7] D. Zuliani,[33,g] and G. Zunica[50]

(LHCb Collaboration)

[1]*School of Physics and Astronomy, Monash University, Melbourne, Australia*
[2]*Centro Brasileiro de Pesquisas Físicas (CBPF), Rio de Janeiro, Brazil*
[3]*Universidade Federal do Rio de Janeiro (UFRJ), Rio de Janeiro, Brazil*
[4]*Department of Engineering Physics, Tsinghua University, Beijing, China*
[5]*Institute Of High Energy Physics (IHEP), Beijing, China*
[6]*School of Physics State Key Laboratory of Nuclear Physics and Technology, Peking University, Beijing, China*
[7]*University of Chinese Academy of Sciences, Beijing, China*
[8]*Institute of Particle Physics, Central China Normal University, Wuhan, Hubei, China*
[9]*Consejo Nacional de Rectores (CONARE), San Jose, Costa Rica*
[10]*Université Savoie Mont Blanc, CNRS, IN2P3-LAPP, Annecy, France*
[11]*Université Clermont Auvergne, CNRS/IN2P3, LPC, Clermont-Ferrand, France*
[12]*Université Paris-Saclay, Centre d'Etudes de Saclay (CEA), IRFU, Saclay, France, Gif-Sur-Yvette, France*
[13]*Aix Marseille Univ, CNRS/IN2P3, CPPM, Marseille, France*
[14]*Université Paris-Saclay, CNRS/IN2P3, IJCLab, Orsay, France*
[15]*Laboratoire Leprince-Ringuet, CNRS/IN2P3, Ecole Polytechnique, Institut Polytechnique de Paris, Palaiseau, France*
[16]*LPNHE, Sorbonne Université, Paris Diderot Sorbonne Paris Cité, CNRS/IN2P3, Paris, France*
[17]*I. Physikalisches Institut, RWTH Aachen University, Aachen, Germany*
[18]*Universität Bonn - Helmholtz-Institut für Strahlen und Kernphysik, Bonn, Germany*
[19]*Fakultät Physik, Technische Universität Dortmund, Dortmund, Germany*
[20]*Physikalisches Institut, Albert-Ludwigs-Universität Freiburg, Freiburg, Germany*
[21]*Max-Planck-Institut für Kernphysik (MPIK), Heidelberg, Germany*
[22]*Physikalisches Institut, Ruprecht-Karls-Universität Heidelberg, Heidelberg, Germany*
[23]*School of Physics, University College Dublin, Dublin, Ireland*
[24]*INFN Sezione di Bari, Bari, Italy*
[25]*INFN Sezione di Bologna, Bologna, Italy*
[26]*INFN Sezione di Ferrara, Ferrara, Italy*
[27]*INFN Sezione di Firenze, Firenze, Italy*
[28]*INFN Laboratori Nazionali di Frascati, Frascati, Italy*
[29]*INFN Sezione di Genova, Genova, Italy*
[30]*INFN Sezione di Milano, Milano, Italy*
[31]*INFN Sezione di Milano-Bicocca, Milano, Italy*
[32]*INFN Sezione di Cagliari, Monserrato, Italy*
[33]*INFN Sezione di Padova, Padova, Italy*
[34]*INFN Sezione di Perugia, Perugia, Italy*
[35]*INFN Sezione di Pisa, Pisa, Italy*
[36]*INFN Sezione di Roma La Sapienza, Roma, Italy*
[37]*INFN Sezione di Roma Tor Vergata, Roma, Italy*
[38]*Nikhef National Institute for Subatomic Physics, Amsterdam, Netherlands*
[39]*Nikhef National Institute for Subatomic Physics and VU University Amsterdam, Amsterdam, Netherlands*
[40]*AGH - University of Krakow, Faculty of Physics and Applied Computer Science, Kraków, Poland*
[41]*Henryk Niewodniczanski Institute of Nuclear Physics Polish Academy of Sciences, Kraków, Poland*
[42]*National Center for Nuclear Research (NCBJ), Warsaw, Poland*
[43]*Horia Hulubei National Institute of Physics and Nuclear Engineering, Bucharest-Magurele, Romania*
[44]*Affiliated with an institute covered by a cooperation agreement with CERN*
[45]*DS4DS, La Salle, Universitat Ramon Llull, Barcelona, Spain*
[46]*ICCUB, Universitat de Barcelona, Barcelona, Spain*
[47]*Instituto Galego de Física de Altas Enerxías (IGFAE), Universidade de Santiago de Compostela, Santiago de Compostela, Spain*
[48]*Instituto de Fisica Corpuscular, Centro Mixto Universidad de Valencia - CSIC, Valencia, Spain*
[49]*European Organization for Nuclear Research (CERN), Geneva, Switzerland*
[50]*Institute of Physics, Ecole Polytechnique Fédérale de Lausanne (EPFL), Lausanne, Switzerland*
[51]*Physik-Institut, Universität Zürich, Zürich, Switzerland*

[52]NSC Kharkiv Institute of Physics and Technology (NSC KIPT), Kharkiv, Ukraine
[53]Institute for Nuclear Research of the National Academy of Sciences (KINR), Kyiv, Ukraine
[54]School of Physics and Astronomy, University of Birmingham, Birmingham, United Kingdom
[55]H.H. Wills Physics Laboratory, University of Bristol, Bristol, United Kingdom
[56]Cavendish Laboratory, University of Cambridge, Cambridge, United Kingdom
[57]Department of Physics, University of Warwick, Coventry, United Kingdom
[58]STFC Rutherford Appleton Laboratory, Didcot, United Kingdom
[59]School of Physics and Astronomy, University of Edinburgh, Edinburgh, United Kingdom
[60]School of Physics and Astronomy, University of Glasgow, Glasgow, United Kingdom
[61]Oliver Lodge Laboratory, University of Liverpool, Liverpool, United Kingdom
[62]Imperial College London, London, United Kingdom
[63]Department of Physics and Astronomy, University of Manchester, Manchester, United Kingdom
[64]Department of Physics, University of Oxford, Oxford, United Kingdom
[65]Massachusetts Institute of Technology, Cambridge, Massachusetts, USA
[66]University of Cincinnati, Cincinnati, Ohio, USA
[67]University of Maryland, College Park, Maryland, USA
[68]Los Alamos National Laboratory (LANL), Los Alamos, New Mexico, USA
[69]Syracuse University, Syracuse, New York, USA
[70]Pontifícia Universidade Católica do Rio de Janeiro (PUC-Rio), Rio de Janeiro, Brazil (associated with Universidade Federal do Rio de Janeiro (UFRJ), Rio de Janeiro, Brazil)
[71]School of Physics and Electronics, Hunan University, Changsha City, China (associated with Institute of Particle Physics, Central China Normal University, Wuhan, Hubei, China)
[72]Guangdong Provincial Key Laboratory of Nuclear Science, Guangdong-Hong Kong Joint Laboratory of Quantum Matter, Institute of Quantum Matter, South China Normal University, Guangzhou, China (associated with Department of Engineering Physics, Tsinghua University, Beijing, China)
[73]Lanzhou University, Lanzhou, China (associated with Institute Of High Energy Physics (IHEP), Beijing, China)
[74]School of Physics and Technology, Wuhan University, Wuhan, China (associated with Department of Engineering Physics, Tsinghua University, Beijing, China)
[75]Departamento de Fisica, Universidad Nacional de Colombia, Bogota, Colombia (associated with LPNHE, Sorbonne Université, Paris Diderot Sorbonne Paris Cité, CNRS/IN2P3, Paris, France)
[76]Ruhr Universitaet Bochum, Fakultaet f. Physik und Astronomie, Bochum, Germany (associated with Fakultät Physik, Technische Universität Dortmund, Dortmund, Germany)
[77]Eotvos Lorand University, Budapest, Hungary (associated with European Organization for Nuclear Research (CERN), Geneva, Switzerland)
[78]Van Swinderen Institute, University of Groningen, Groningen, Netherlands (associated with Nikhef National Institute for Subatomic Physics, Amsterdam, Netherlands)
[79]Universiteit Maastricht, Maastricht, Netherlands (associated with Nikhef National Institute for Subatomic Physics, Amsterdam, Netherlands)
[80]Tadeusz Kosciuszko Cracow University of Technology, Cracow, Poland (associated with Henryk Niewodniczanski Institute of Nuclear Physics Polish Academy of Sciences, Kraków, Poland)
[81]Universidade da Coruña, A Coruña, Spain (associated with DS4DS, La Salle, Universitat Ramon Llull, Barcelona, Spain)
[82]Department of Physics and Astronomy, Uppsala University, Uppsala, Sweden (associated with School of Physics and Astronomy, University of Glasgow, Glasgow, United Kingdom)
[83]University of Michigan, Ann Arbor, Michigan, USA (associated with Syracuse University, Syracuse, New York, USA)

[a]Deceased.
[b]Also at Lamarr Institute for Machine Learning and Artificial Intelligence, Dortmund, Germany.
[c]Also at Università degli Studi di Milano-Bicocca, Milano, Italy.
[d]Also at Università di Roma Tor Vergata, Roma, Italy.

[e]Also at Scuola Normale Superiore, Pisa, Italy.
[f]Also at Università di Ferrara, Ferrara, Italy.
[g]Also at Università di Padova, Padova, Italy.
[h]Also at Facultad de Ciencias Fisicas, Madrid, Spain.
[i]Also at Università di Bologna, Bologna, Italy.
[j]Also at Università di Genova, Genova, Italy.
[k]Also at Università degli Studi di Milano, Milano, Italy.
[l]Also at Universidad Nacional Autónoma de Honduras, Tegucigalpa, Honduras.
[m]Also at Università di Cagliari, Cagliari, Italy.
[n]Also at Centro Federal de Educacão Tecnológica Celso Suckow da Fonseca, Rio De Janeiro, Brazil.
[o]Also at Università di Bari, Bari, Italy.
[p]Also at Center for High Energy Physics, Tsinghua University, Beijing, China.
[q]Also at Università di Perugia, Perugia, Italy.
[r]Also at LIP6, Sorbonne Université, Paris, France.
[s]Also at Università di Pisa, Pisa, Italy.
[t]Also at Hangzhou Institute for Advanced Study, UCAS, Hangzhou, China.
[u]Also at School of Physics and Electronics, Henan University, Kaifeng, China.
[v]Also at Università di Bergamo, Bergamo, Italy.
[w]Also at Università di Siena, Siena, Italy.
[x]Also at Department of Physics/Division of Particle Physics, Lund, Sweden.
[y]Also at Università della Basilicata, Potenza, Italy.
[z]Also at Universidad de Alcalá, Alcalá de Henares, Spain.
[aa]Also at Università di Urbino, Urbino, Italy.